# Local strain engineering in atomically thin MoS$_2$

*Andres Castellanos-Gomez* [1,*], *Rafael Roldán* [2,*], *Emmanuele Cappelluti* [2,3], *Michele Buscema* [1], *Francisco Guinea* [2], *Herre S. J. van der Zant* [1] *and Gary A. Steele* [1].

[1] Kavli Institute of Nanoscience, Delft University of Technology, Lorentzweg 1, 2628 CJ Delft, The Netherlands.
[2] Instituto de Ciencia de Materiales de Madrid, CSIC, Sor Juana Ines de la Cruz 3, 28049 Madrid, Spain.
[3] Institute for Complex Systems (ISC), CNR, U.O.S. Sapienza, v. dei Taurini 19, 00185 Rome, Italy.
a.castellanosgomez@tudelft.nl , rroldan@icmm.csic.es

Controlling the bandstructure through local-strain engineering is an exciting avenue for tailoring optoelectronic properties of materials at the nanoscale. Atomically thin materials are particularly well suited for this purpose because they can withstand extreme non-homogeneous deformations before rupture. Here, we study the effect of large localized strain in the electronic bandstructure of atomically thin MoS$_2$. Using photoluminescence imaging, we observe a strain-induced reduction of the direct bandgap, and funneling of photogenerated excitons towards regions of higher strain. To understand these results, we develop a non-uniform tight-binding model to calculate the electronic properties of MoS$_2$ nanolayers with complex and realistic local strain geometries, finding good agreement with our experimental results.

Tuning the band structure of a material by subjecting it to strain constitutes an important strategy to enhance the performance of electronic devices.[1] Using local strain, confinement potentials for excitons can be engineered, with possibilities for trapping excitons for quantum optics [2] and for efficient collection of solar energy.[3, 4] Two-dimensional materials are able to withstand large strains before rupture,[5-7] offering a unique opportunity to introduce large local strains. Here, we study atomically thin MoS$_2$ layers [8-13] with large local strains of up to 2.5% induced by controlled delamination from a substrate. Using simultaneous scanning Raman and photoluminescence imaging, we spatially resolve a direct bandgap reduction of up to 90 meV induced by local strain. We observe a funnel effect in which excitons drift hundreds of nanometers to lower bandgap regions before recombining, demonstrating exciton confinement by local strain. The observations are supported by an atomistic tight-binding model developed to predict the effect of inhomogeneous strain on the local electronic states in MoS$_2$. The possibility of generating large strain-induced variations in exciton trapping potentials opens the door for a variety of applications in atomically thin materials including photovoltaics, quantum optics and two-dimensional optoelectronic devices.

Atomically thin MoS$_2$ is a semiconducting analogue to graphene, with a large intrinsic bandgap [9] and large Seebeck coefficient,[14] that can sustain elastic deformations up to 25% before the rupture.[6, 7, 15] This large breaking strength value has motivated a surge of theoretical works that study the changes of the MoS$_2$ band structure induced by strain [16-25] suggesting a means to trap photogenerated excitons.[3] In a conventional semiconductor, exciton confining potentials are typically engineered by locally changing its chemical composition. Alternatively, techniques such as laser trapping of excitons in a single material by the AC Stark effect has been employed to create traps with a 5 meV confining potential –a value suited for quantum optics experiments.[2] The large rupture strength of 2D crystals allows one to induce large local strains by bending or folding the material like a piece of paper. In this letter, we intentionally wrinkle few-layer MoS$_2$ flakes to subject them to large, local uniaxial strain. The applied strain is quantified by a combination of atomic force microscopy and Raman spectroscopy and the effect of the non-uniform strain on the bandgap is spatially resolved by scanning photoluminescence.

Large localized uniaxial strain (up to 2.5% tensile) in few-layers MoS$_2$ samples (3 to 5 layers) has been achieved in the following way: first MoS$_2$ flakes are deposited onto an elastomeric substrate (see Materials and Methods in the Supplementary Information) which is pre-stretched by 100%. Subsequently, the tension in the elastomeric substrate is suddenly released generating well-aligned wrinkles in the MoS$_2$ layers perpendicular to the initial uniaxial strain axis in the substrate (see Fig. 1a). The mechanism behind the formation of these wrinkles is buckling-induced delamination.[26] We have found that this fabrication procedure reproducibly generates large-scale wrinkles (microns in height, separated by tens of microns) for bulk MoS$_2$ flakes (more than 10-15 layers) while thin MoS$_2$ layers (3-5 layers) exhibit wrinkles that are between 50 nm and 350 nm in height and are separated by few microns. Figure 1b and 1c show an optical and a topographic atomic force microscopy image of a wrinkled MoS$_2$ flake with regions with different number of layers. Wrinkles in single and bilayer MoS$_2$ samples are not stable and they tend to collapse forming folds (see Supplementary Information).

The maximum uniaxial tensile strain $\varepsilon$ is accumulated on top of the winkles and can be estimated as [26]

$$\varepsilon \sim \pi^2 h\, \delta\, /\, (1-\sigma^2)\lambda^2\ , \quad [1]$$

where $\sigma$ is the MoS$_2$ Poisson's ratio (0.125), $h$ is the thickness of the flake and $\delta$ and $\lambda$ are the height and width of the wrinkle respectively. The values for $\delta$ and $\lambda$ are extracted from the atomic force microscopy characterization of the wrinkle geometry (see Figure 1c). To accurately determine the thickness, we employed a combination of atomic force microscopy, quantitative optical microscopy,[27] Raman spectroscopy [21,22] and photoluminescence [23,24] (see Supplementary Information). For the wrinkled thin MoS$_2$ flakes (3 to 5 layers), the estimated uniaxial strain ranges from 0.2% to 2.5%. Interestingly, despite the large strain values, the wrinkles are stable in time and no slippage has been found. Note that slippage is usually a limiting factor in experiments applying uniform uniaxial strain to atomically thin crystals (graphene, MoS$_2$) in bending geometries.[28,29]

Raman spectroscopy has proven to be a powerful tool to characterize graphene samples subjected to uniaxial strain.[28] Here we employ the same technique to study the changes of the vibrational modes of MoS$_2$ flakes induced by localized strain. Figure 2a shows the Raman spectra measured on the flat region and on top of a wrinkle for a four layers thick MoS$_2$ flake (blue and red traces respectively). We observe that on top of the wrinkle the two most prominent Raman peaks, the $E^1_{2g}$ and the $A_{1g}$,[30,31] are red-shifted (the vibrations soften). The $A_{1g}$ mode (that corresponds to the sulfur atoms oscillating in anti-phase out-of-plane) is less affected than the $E^1_{2g}$ mode (sulfur and molybdenum atoms oscillating in anti-phase parallel to the crystal plane). It has been shown that the $A_{1g}$ peak shift can be used to estimate the doping level in atomically thin MoS$_2$ [32,33]. However, a change in doping level enough to noticeably modify the optoelectronic properties of MoS$_2$ would yield a shift of the $A_{1g}$ peak much larger than the one we measure [32,33]. In fact, the slight red-shift of the $A_{1g}$ peak observed on the wrinkled MoS$_2$ and the larger red-shift of the $E^1_{2g}$ peak are in good agreement with a recent study on the effect of uniform tensile uniaxial strain on the Raman spectrum of MoS$_2$.[29] In fact, for thin plates subjected to large deformations (much larger than their thickness) the neutral axis lays out of the plate yielding a net tensile strain, in agreement to the Raman measurements on wrinkled MoS$_2$.[34, 35]

To study the effect of localized tensile strain on the band structure of few-layer MoS$_2$, we carried out scanning photoluminescence measurements on the wrinkled samples. Although few-layer MoS$_2$ is an indirect bandgap semiconductor, its photoluminescence spectrum is dominated by the direct gap transitions, at the K point of the Brillouin zone, between the valence band (which is split by interlayer and spin-orbit coupling) and the conduction band.[36, 37] These direct gap transitions show up in the photoluminescence spectra as two resonances, known as the A and B excitons (Figure 2b).[36, 37] The indirect bandgap transition also contributes to the photoluminescence spectra, showing off as a very weak peak at lower energy that the A and B excitons (see the Supporting Information for more details on the strain tuning of the indirect bandgap). Interestingly, when the photoluminescence is acquired on top of a wrinkle, both the A and B excitons are red-shifted with respect to the spectrum measured on a flat region of the same flake. This indicates that the uniaxial strain localized on the top of the wrinkle modifies the band structure around the K point, reducing the energy of the direct band gap transition. Note that the bandgap determined by photoluminescence spectroscopy (usually referred to as optical bandgap) differs from that determined by electronic transport due to the exciton binding energy.[3, 38] Recent theoretical works have estimated that the exciton binding energy in MoS$_2$ (3 to 5 layers thick) is on the order of 100 meV.[39] While this value is indeed large, it should not be significantly influenced by the strain we apply, as was recently demonstrated theoretically in reference [3] that the exciton binding energy shifts by less than 6 meV per % of strain. As these corrections are small compared to the shifts of the photoluminescence peak we report, we can neglect their expected small contribution, and thus imply the changes in bandstructure of MoS$_2$ from the observed changes in the photoluminescence spectra. We have spatially resolved the changes in the vibrational modes and the band structure induced by this non-uniform strain by performing simultaneous scanning Raman microscopy and photoluminescence (see for an example Figure 2c-e): Figure 2d and 2e show how the shifts of the vibrational modes and of the exciton wavelengths occur at the wrinkles location. This observation underlines the potential of non-uniform strain to engineer a spatial variation of the vibrational and optoelectronic properties of atomically thin MoS$_2$ crystals.

As in the case of graphene, the shift on the Raman modes upon applied strain can be employed to quantify the strain load on the sample.[28] Figure 3a shows the correlation between the shift of the $E^1_{2g}$ Raman mode and the strain, estimated from Eq. [1] using the AFM topography measured for more than 50 wrinkles with thickness ranging from 3 to 5 layers. We find that the $E^1_{2g}$ Raman peak shifts by -1.7 cm$^{-1}$ per % strain, in agreement with recent results obtained by Rice *et al.* applying uniform uniaxial strain to MoS$_2$.[29] Note that the dispersion in the estimated strain is due to the systematic uncertainties in the determination of $h$, $\delta$ and $\lambda$.

Using this quantitative *in-situ* calibration of the strain based on the Raman spectrum, we can directly relate the change in the direct bandgap transition energy to the strain induced in wrinkles in few-layer MoS$_2$ flakes. Figure 3b shows the results obtained on more than 50 different wrinkles (3 to 5 layers in thickness). The direct gap transition energy decreases for increasing uniaxial strain values; for a ~2.5 % tensile strain, the change is about -90 meV, which corresponds to a reduction of the direct bandgap transition energy of 5 %. The change in the direct bandgap of MoS$_2$ induced by the non-uniform strain generated by the wrinkles is comparable to that achieved in semiconducting nanowires by using a straining dielectric envelope,[40] and about 5 times larger than the change induced in quantum dots by biaxial straining with piezoelectric actuators.[41] Nonetheless, unlike these previous strain engineering approaches, our procedure using non-uniform strain allows us to locally modify the band structure of semiconductors on the nanometer scale.

In order to gain further insight into the band structure change induced by the non-uniform strain, we have developed a description of the electronic properties of MoS$_2$ based on a tight-binding (TB) model.[42] Such an approach is particularly advantageous to investigate and to model systems under inhomogeneous conditions (e.g. strain or electrostatic potential) where the required supercell size makes the computational analysis unaffordable with other methods such as density functional theory. The TB description also provides a suitable basis for further inclusion of many-body electron-electron effects as well as of the dynamical effects of the electron-lattice interaction. On the microscopic level, the description in terms of tight-binding parameters permits also to investigate separately the role of each interatomic ligand on the band structure, both at the uniform and at the local level.

The TB model is expressed in terms of few interatomic Slater-Koster orbital ligands $V_{\mu,\nu}(r_\alpha - r_\nu)$, where $r_\mu$ denotes the position of the atoms labeled by the quantum number $\mu$, which includes the orbital and the cell basis degree of freedoms. Inhomogeneous strain effects are included by changing the respective atomic positions $r_\mu$ according to the local strain tensor,[43,44] which modulates the TB parameters through the electron-phonon coupling $\beta_{\mu,\nu} = -d \log V_{\mu,\nu}(R)/d \log R$, where $R$ is the interatomic distance. In the absence of a microscopic estimate of the hopping-resolved $\beta_{\mu,\nu}$, we assume $\beta_{\mu,\nu} = 3.2$, similar to the case of graphene.[45]

With this theoretical model, we are able to simulate different realistic local strain configurations. Figure 4 shows the calculated band structure for zigzag MoS$_2$ ribbons (infinite along the *x* axis and 200 unit cells wide along the *y* axis) subjected to the local strain induced by a wrinkle (see insets in Figure 4). The finite width of the MoS$_2$ ribbons leads to the formation of electronic bands composed by the accumulation of *N* sub-bands (with *N* the number of unit cells along the width of the ribbon). Arrows indicate the direct band gap transitions and from the figure one clearly sees that the corresponding band-gap energy decreases as the uniaxial strain increases. An analysis of the local density of states of wrinkled MoS$_2$ ribbons (see section 10.5 of the Supp. Info.) reveals that the minimum value of the bandgap (indicated with vertical arrows in Figure 4) is located in the middle of the wrinkle where the strain value is maximum.

By repeating the simulations presented in Figure 4 for wrinkled ribbons (with *N* = 1200 unit cells) with different amounts of strain, we find that that the direct bandgap decreases linearly upon the maximum uniaxial strain value (solid line in Figure 3b). Examining Figure 3b, we conclude that the theory describes well the extremal points of the distribution; however, there are also many points that fall below the solid line. The fact that there are many points that fall below the line is a result of a combination of the finite size of the local laser probe we are using, together with the local nature of the strain in the sample. In particular, the width λ of the wrinkles (between 700 nm and 1200 nm) is comparable with the size of the laser spot (400 nm). In this situation the optical measurements cannot be considered purely local but they rather probe a finite portion of the wrinkle; this affects the Raman and photoluminescence measurements in a different way. In the Raman process, the photon emitted in a Raman scattering event is from the same position where it was absorbed.

Photoinduced excitons in contrast move to lower bandgap regions before recombining. The non-uniform bandgap profile induced by the local strain of the wrinkles therefore generates a trap for photogenerated excitons with a depth of up to 90 meV. The result, referred to as the funnel effect [3] (see section 10 of the Supp. Info for more details), means that while the Raman signal represents an average over the whole laser spot area, the photoluminescence gives only information about the minimum gap in the region illuminated by the laser (Figure 3c). When including this funnel effect, the local strain tight binding model fits very well with our experimental results (dashed lines in Figure 3b). The agreement of the data with the local strain model is also better than that with a model that contains only uniform strain, which tends to underestimate the bandgap shift (see Supplementary Information).

In summary, we have demonstrated tuning of the electronic band structure of atomically thin layers of MoS$_2$ by engineering local strain. A tight-binding model including this non-uniform strain was developed to understand changes of the electronic band structure at the local scale, and is found to be in good agreement with our experimental observations. The capability to engineer a local confinement potential for excitons using strain provides a unique means to design and control the optoelectronic properties of atomically thin MoS$_2$-based devices. The technique we present here offers a route to local strain engineering in both MoS$_2$ and other atomically thin crystals, opening up many applications in diverse fields such as optics, electronics, optoelectronics, photovoltaics and surface science.

During preparation of our manuscript, we became aware recent works studying the relation between bandgap and strain in MoS$_2$ flakes employing uniform uniaxial strain in a conventional three point bending configuration[46] or uniform biaxial strain using piezoelectric substrates, but which did not address the possibility of local strain engineering in thin 2D crystals.

SUPPORTING INFORMATION

Supporting information file can be downloaded from:

http://pubs.acs.org/doi/suppl/10.1021/nl402875m/suppl_file/nl402875m_si_001.pdf

ACKNOWLEDGMENT

The authors acknowledge the fruitful discussions with Lihao Han and Arno Smets about the photoluminescence setup. This work was supported by the European Union (FP7) through the program RODIN and the Dutch organization for Fundamental Research on Matter (FOM). A.C-G. acknowledges financial support through the FP7-Marie Curie Project PIEF-GA-2011-300802 ('STRENGTHNANO'). R.R. acknowledges financial support from the Juan de la Cierva Program (MINECO, Spain). E.C. acknowledges financial support through the FP7-Marie Curie Project PIEF-GA-2009-251904. F.G. and R. R. acknowledge support from MINECO (Spain), through grant FIS2011- 23713, and ERC Advanced Grant #290846.

FIGURES

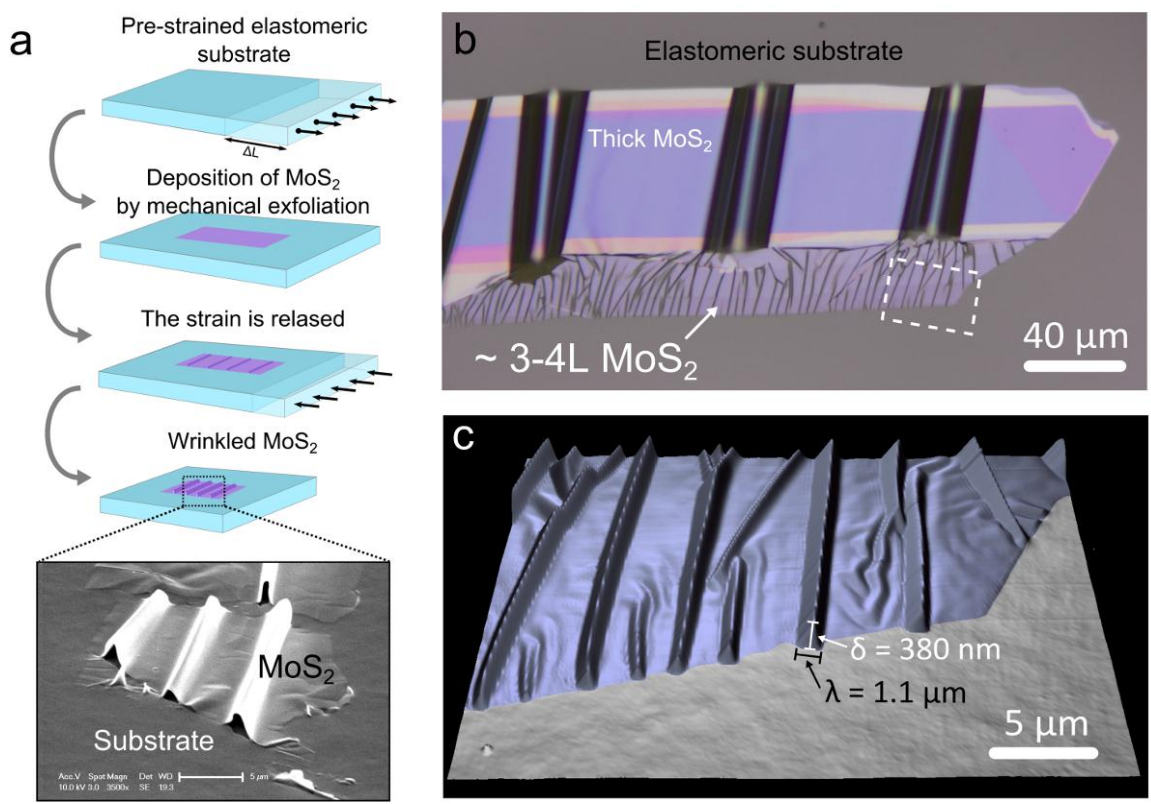

**Figure 1. Localized uniaxial strain in MoS$_2$.** (a) Schematic diagram of the fabrication process of wrinkled MoS$_2$ nanolayers. An elastomeric substrate is stretched prior depositing MoS$_2$ by mechanical exfoliation. The strain is released afterwards producing buckling-induced delamination of the MoS$_2$ flakes. (b) Optical

microscopy image of a wrinkled MoS$_2$ flake fabricated by buckling-induced delamination. (c) Atomic force microscopy (AFM) topography image of the region marked by the dashed rectangle in (b).

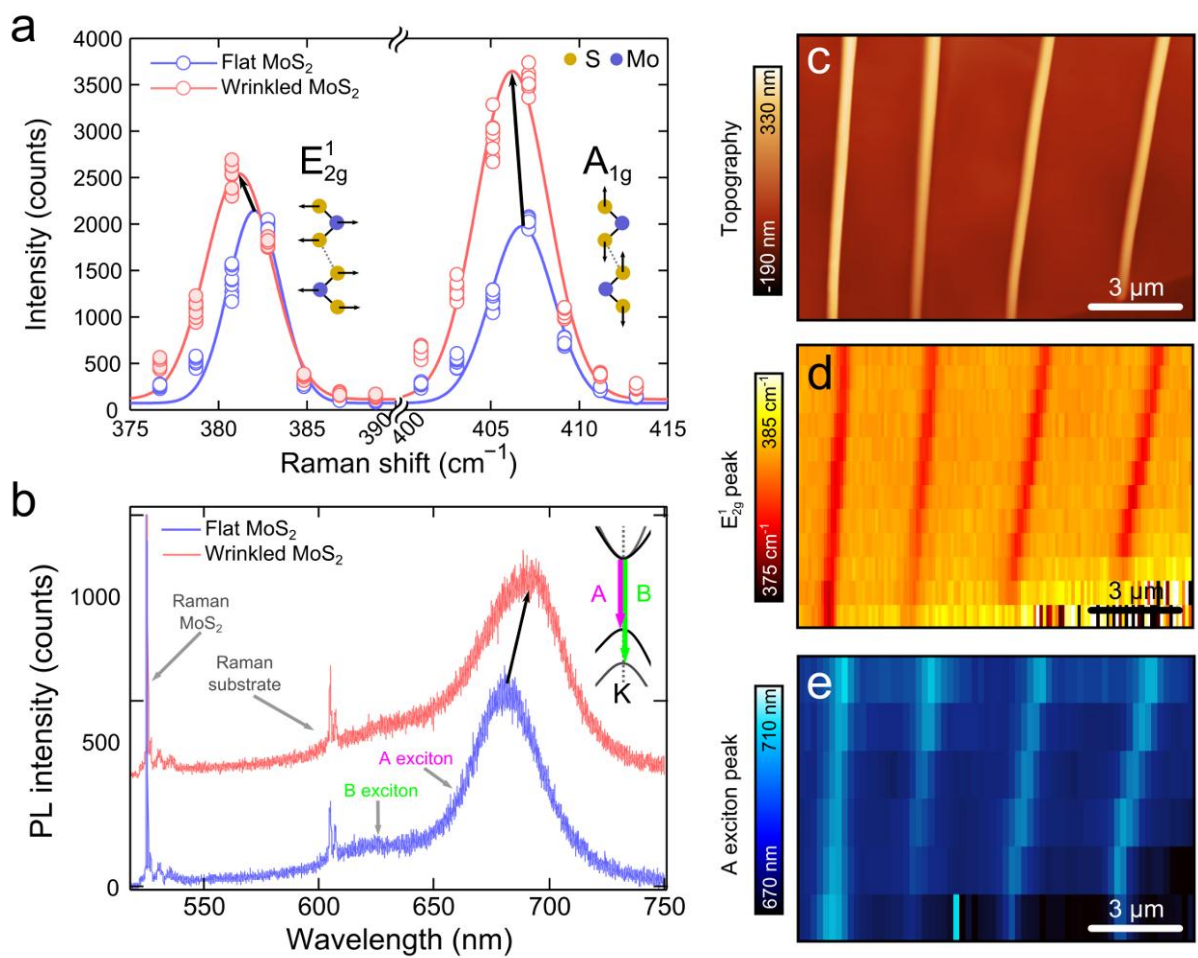

**Figure 2. Raman and Photoluminiscence spectra of strained MoS$_2$.** (a) Raman spectra measured on a flat (blue) and on a wrinkled (red) region of a 4 layers thick MoS$_2$ flake. Although both the E$^1_{2g}$ and the A$_{1g}$ modes are shifted towards lower Raman shift, the E$^1_{2g}$ mode presents the higher shift. (b) Photoluminescence spectra measured on the flat region (blue) and on top of the wrinkle (red) in the same MoS$_2$ flake. Notice that the red spectrum has been vertically shifted for clarity. The photoluminescence emission from the wrinkle is red shifted with respect to the one of the flat MoS$_2$. (inset in (b)) Schematic diagram of the direct transitions (A and B exciton) at the K point, observed in the photoluminescence spectra. (c) AFM topography image of a 4L thick MoS$_2$ flake with 4 wrinkles. (d) and (e) show the Raman shift of the E$^1_{2g}$ vibrational mode and the wavelength of the A exciton peak respectively, measured in the same region as (c).

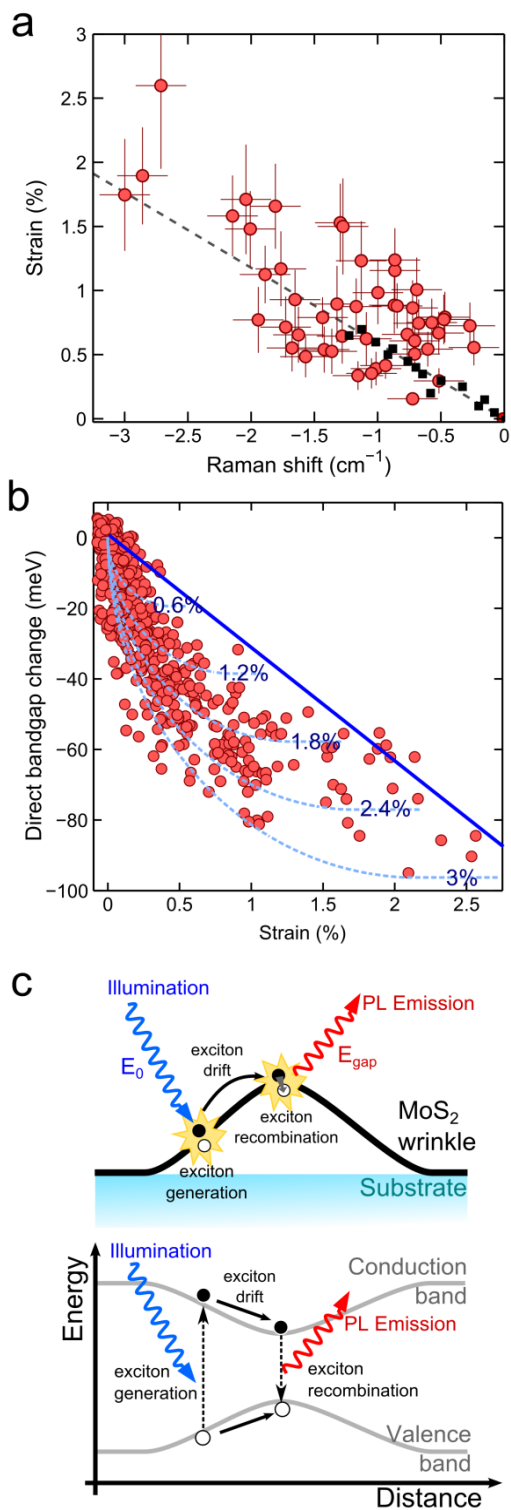

**Figure 3. Strain tuning the direct bandgap transitions of MoS$_2$.** (a) Correlation between the measured Raman shift of the $E^1_{2g}$ mode and the strain estimated from the analysis of the AFM topography of more than 50 wrinkles with thickness ranging from 3 to 5 layers (red circles). The black squares show the results reported in Ref. [29] (which follows a linear relationship with a slope -1.7 cm$^{-1}$ per % strain) obtained by applying uniform tensile strain to few-layer MoS$_2$. (b) Change in the energy of the direct bandgap transition as a function of the strain measured by scanning the laser across more than 50 wrinkles. The change in the direct bandgap is obtained from the shift of the A exciton in the photoluminescence spectra while the strain can be estimated from the $E^1_{2g}$ Raman mode shift. The solid line in (b) is the bandgap *vs.* strain relationship calculated for wrinkled MoS$_2$ ribbons with different levels of maximum strain employing the TB model discussed in the text. The dashed lines show the expected bandgap *vs.* strain relationship after accounting for the effect of the finite laser spot size and the funnel effect. (c) Schematic diagram explaining the funnel effect due to the non-homogeneous strain in the wrinkled MoS$_2$.

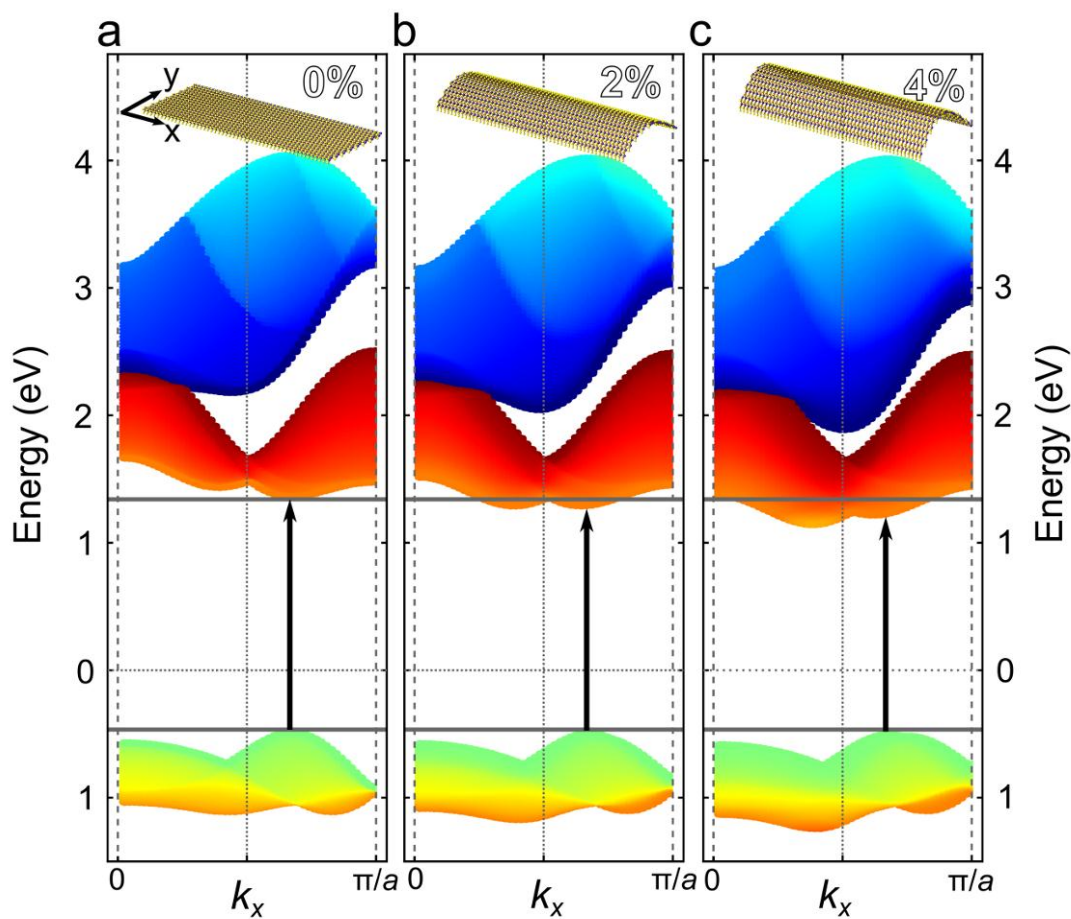

**Figure 4. Band structure for non-uniform strained MoS$_2$.** Calculated electronic band structure for a zigzag MoS$_2$ ribbon (infinite along the *x* axis) with periodic boundary conditions and 200 unit cells wide along *y* axis (*a* is the unit cell size *a* = 3.16 Å). Each band is composed by 200 sub-bands, that arise from the discrete width of the ribbon in the y-direction. These sub-bands are plotted here as a set of lines, densely overlapping, where each sub-band is plotted with a line of different color. (a) The band structure for the case of an unstrained ribbon. (b), (c) Band structures calculated for wrinkled ribbons with a 2% and 4% maximum tensile strain respectively. In the photoluminescence experiments, the wavelength of the emitted light is determined by the direct bandgap transition energy, indicated in each panel by a vertical arrow.